\documentclass{aastex}
\usepackage{spr-astr-addons}
\usepackage{url}

\RequirePackage{color}
\usepackage{graphicx}
\usepackage[latin1]{inputenc}
\usepackage{amsmath}

\usepackage{gensymb}       

\usepackage{hyperref}

\usepackage{colortbl}
\usepackage{pstricks,pst-grad}
\usepackage{pst-pdf}

\usepackage{fancybox, graphicx}
\graphicspath{{./}{./figures/}}

\definecolor{lightBrown}{rgb}{0.9,0.8,0.6}
\definecolor{lightGreen}{rgb}{0.8,1.0,0.8}
\definecolor{lightYellow}{rgb}{1.0,1.0,0.8}
\definecolor{lightBlue}{rgb}{0.0,0.4,0.8}
\definecolor{lightRed}{rgb}{1.0,0.8,0.8}
\definecolor{darkYellow}{rgb}{0.9,0.7,0.1}
\definecolor{titleBlue}{rgb}{0.0,0.2,0.6}

\newrgbcolor{white}{1.0 1.0 1.0}
\newrgbcolor{lightGreen}{0.8 1.0 0.8}
\newrgbcolor{lightYellow}{1.0 1.0 0.8}
\newrgbcolor{lightGrey}{0.6 0.6 0.6}
\newrgbcolor{lightBlue}{0.0 0.4 0.8}
\newrgbcolor{mediumRed}{0.4 0.1 0.1}
\newrgbcolor{darkCyan}{0.25 0.5 0.5}
\newrgbcolor{darkGrey}{0.4 0.4 0.4}
\newrgbcolor{darkBrown}{0.3 0.2 0.1}
\newrgbcolor{darkBlue}{0.0 0.1 0.3}
\newrgbcolor{darkGreen}{0.1 0.3 0.1}
\newrgbcolor{darkOrange}{0.8 0.4 0.0}
\newrgbcolor{darkRed}{0.6 0.2 0.2}
\newrgbcolor{darkViolet}{0.2 0.0 0.2}
\newrgbcolor{darkYellow}{0.9 0.7 0.1}
\newrgbcolor{veryDarkCyan}{0.20 0.4 0.4}

\newlength{\parboxWidth}          

\newcommand{\myShadowBox}[1]{
\setlength{\fboxrule}{2pt}
\setlength{\shadowsize}{3pt}
\settowidth{\parboxWidth}{#1}
\addtolength{\parboxWidth}{4ex}
\shadowbox{\parbox{\parboxWidth}{\begin{center} \black #1 \end{center}
\vspace{-1ex}}}
}

\begin{document}
\title{Developing Ecospheres on Transiently Habitable Planets:\\  
       The Genesis Project}
\shorttitle{The Genesis Project}
\author{Claudius Gros\altaffilmark{1}}
\affil{Institute for Theoretical Physics, Goethe
University Frankfurt, Germany} 


\begin{abstract}
It is often presumed, that life evolves relatively fast 
on planets with clement conditions, at least in its 
basic forms, and that extended periods of habitability 
are subsequently needed for the evolution of higher 
life forms. Many planets are however expected to be 
only transiently habitable. On a large set of 
otherwise suitable planets life will therefore just not 
have the time to develop on its own to a complexity level 
as it did arise on earth with the cambrian explosion. The 
equivalent of a cambrian explosion may however have the 
chance to unfold on transiently habitable planets if it 
would be possible to fast forward evolution by 
3-4 billion years (with respect to terrestrial timescales).
We argue here, that this is indeed possible when seeding the 
candidate planet with the microbial lifeforms, bacteria and 
unicellular eukaryotes alike, characterizing earth before the 
cambrian explosion. An interstellar mission of this kind, 
denoted the `Genesis project', could be carried out by a
relatively low-cost robotic microcraft equipped with a on-board 
gene laboratory for the in situ synthesis of the microbes.

We review here our current understanding of the processes
determining the timescales shaping the geo-evolution of an 
earth-like planet, the prospect of finding Genesis candidate 
planets and selected issues regarding the mission layout. 
Discussing the ethical aspects connected with a Genesis mission, which 
would be expressively not for human benefit, we will also
touch the risk that a biosphere incompatibility may arise
in the wake of an eventual manned exploration of a second earth.

\end{abstract}

\keywords{astrobiology, habitable planets, interstellar missions}


\section{Introduction}

Three ongoing lines of research have progressed in the last
years to a point which allows us to assess now the feasibility 
of sending out interstellar probes with the mission of
bringing life to otherwise barren exoplanets.

In first place comes here the insight from exoplanet 
search efforts, that the diversity of the hitherto 
discovered exoplanetary systems is very high. This 
implies, in particular, that no two habitable planets 
may be alike \citep{gudel2014astrophysical} and that 
there will be many planets having only limited periods 
of habitability. There may hence exist in our galaxy a 
plethora of planets where life could truly thrive, having 
however not enough time to fully develop on its own. The key 
idea of the Genesis project is to bring life to these
kind of exoplanets. 

The Genesis project is based furthermore on the evolving consensus
\citep{long2011deep,projectDragonfly,breakthroughinitiatives}, that 
robotic interstellar missions may be realizable within a foreseeable 
future. A conceivable scenario would be in this context to 
accelerate lightweight interstellar probes with ground- or 
orbit-based arrays of powerful lasers
\citep{zhang2015orbital,brashears2015directed}. Decelerating
could then be achieved, on arrival, using magnetic and/or
electric sails \citep{zubrin1991magnetic,perakis2016combining}.

The progress \citep{gibson2010creation,hutchison2016design} 
achieved recently in creating synthesized and minimal 
(in terms of the genome) cells, indicates furthermore that 
humanity will acquire most probably already within a few 
decades the capability to synthesize a vast palette of life 
forms from scratch. We can hence envision that a Genesis 
probe would be able to cultivate in situ many different 
types of microbes using a robotic gene laboratory.

The Genesis project consists hence of three steps.
\begin{itemize}
\item[--] Searching for transiently habitable planets.
\item[--] Sending interstellar robotic crafts for detailed investigations.
\item[--] Seeding the candidate planet with in situ synthesized lifeforms.
\end{itemize}
In this study we will review the prospects of discovering
Genesis candidate planets, the time scale for evolutionary
speedup one may realistically hope to achieve and the time scales 
the Genesis process may take to unfold. The Genesis project
comes of course with serious ethical caveats regarding
in particular planetary protection, which we will also
discuss.

The Genesis project is in first place not for human benefit. 
The key idea is to initiate a self-sustained evolutionary 
process, which will then carry the developing biosphere of
the host planet to its own future. Later stage human interventions 
are not excluded, but not necessary. The same holds for an eventual 
human settlement of the candidate planet. In this context we
will also discuss the prospective that biosphere incompatibilities 
may accompany manned missions to second-earth like
exoplanets.

\begin{figure*}[t]
\centering
\includegraphics[width=0.90\textwidth,angle=0]{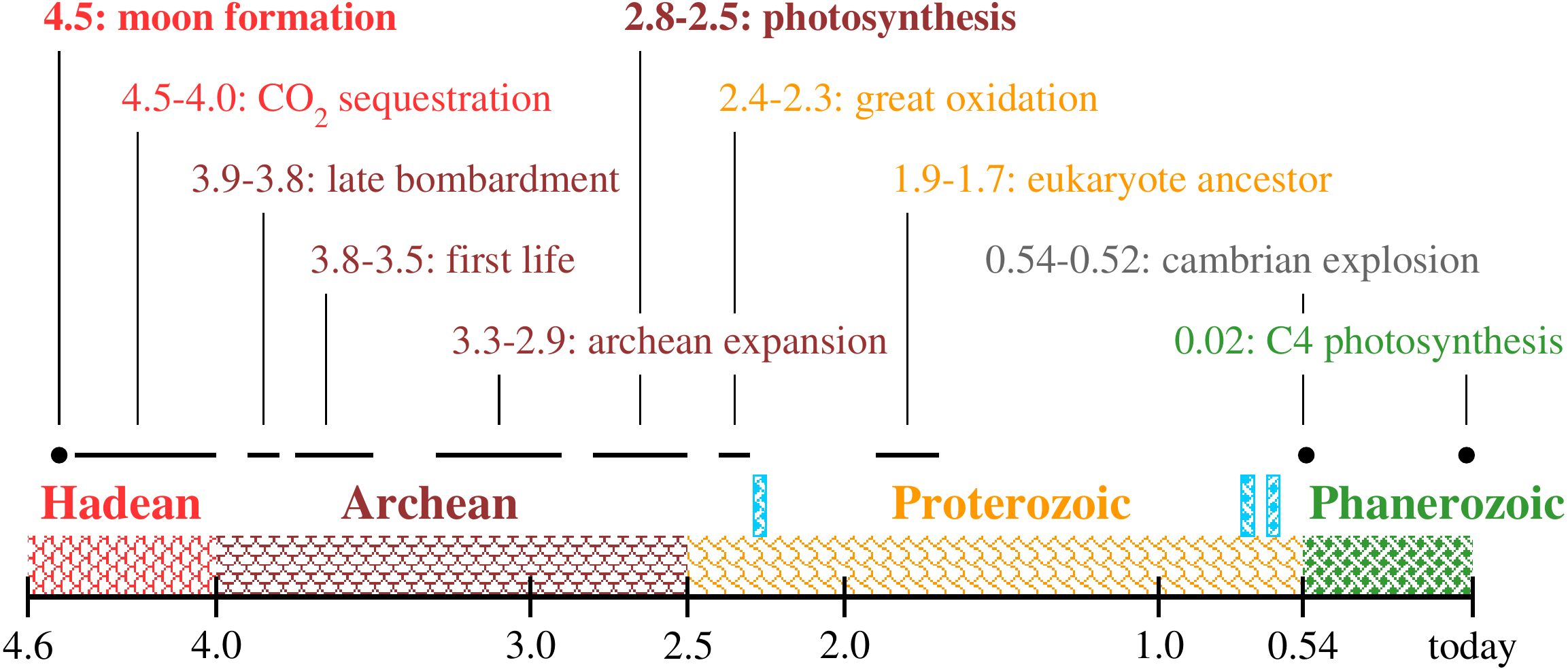}
\caption{Some major milestones in the history of our home planet
         (time in $10^9$-years, Ga). A one-shot Genesis probe may
         achieve an evolutionary fast forward for the candidate
         planet to eukaryotic life, viz by about three Ga.
         The blue bars denote episodes of global glaciation
         (snowball earth). Huronian, Sturtian and Marinoan
         at 2.27\,Ga and 716.5/635\,Ma 
         \citep{tang2013global,macdonald2010calibrating}.
        }
\label{fig:earthHistory}
\end{figure*}

\subsection{Genesis candidate planets}

We currently do not know whether the 
bio-geological evolution of earth has been 
fast or slow in relation to evolutionary
processes on other habitable planets. We 
know however, that the timescale of major 
geo-evolutionary steps has been typically
of the order of one or more billion years 
(Ga). Taking hence one Ga as a reference time 
we may consider a planet to be transiently
habitable if the time span it remains in the 
habitable zone (HZ) ranges from at least
a few hundred million years to about 1-2\,Ga. 

There are various conceivable scenaria why
a planet may be habitable for a finite, 
but prolonged period.

\begin{itemize}
\item[--] {\bf Shift of the habitable zone.}\newline
      Main sequence stars as our sun, which was initially 
      about 30\% less luminous, become brighter with 
      the eons passing. A planet starting out at inner 
      edge of the HZ may hence become eventually too hot 
      \citep{rushby2013habitable}. An initially too far away 
      planet may reversely warm up, but possibly only after 
      a few Ga have passed. In the later case not enough 
      time for the unfolding of a full-fledged evolutionary 
      process may be left.

\item[--] {\bf Long-term orbital instabilities.}\newline
      Several kinds of long-term orbital instabilities, 
      like the Hill- and the Lagrange instability discussed 
      in Sect.~\ref{sect:hill_instability}
      and \ref{sect:lagrange_instability}, may throw a planet 
      out of the habitable zone or, reversely, promote a planet 
      into the HZ \citep{jones2006habitability}. Habitability
      may also be interrupted by later stage orbital resonances, 
      as the one which has possibly caused the late heavy 
      bombardment of our home planet (compare Sect.~\ref{sec:LTB}).

\item[--] {\bf Indigenous processes.}\newline
      An initially habitable planet may become inhabitable 
      also all by itself. Various indigenous processes are
      conceivable in this context.
      \begin{itemize} 
      \item Plate tectonic may cease functioning after 
            a certain time, giving way to the type of
            stagnant lid magma convection presumable
            in place on Venus. Episodic overturns of 
            the crust may then result in global resurfacing
            events.
      \item Shifts in the CO$_2$ balance could lead to the
            complete depletion of atmospheric CO$_2$ levels. 
            On earth this is actually happening, as discussed
            in Sect.~\ref{sec:C4}, albeit only very slowly. 
            A continuous accumulation of CO$_2$ could result 
            on the other side in either a catastrophic runaway, 
            or in a welcome Greenhouse effect.
      \end{itemize}

\end{itemize}

Reviewing the prospects of discovering Genesis candidate we
will rely in part on an analysis of the timeline of the 
geo-evolutionary history of our home planet. Of particular 
interest for the Genesis project are here the involved time 
scales and the question, whether alternative evolutionary 
routes would have been possible.

\section{Terrestrial timescales}

A central question of the present study regards 
the time one may expect a Genesis process will
need to unfold on an exoplanet. As a backdrop to
this question we start with a short review of some 
of the key events in the geo-biological
evolution of earth, where a self-organized
Genesis process is known to have occurred. Times 
will be given either in Giga-years ($10^9$, Ga) 
or Mega-years ($10^6$, Ma). Some of the key events 
shaping our home planet are shown on scale 
in Fig.~\ref{fig:earthHistory}. 

Shortly after earth was formed together with
most of the solar system about 4.6\,Ga ago, an
impact with a Mars-size object led to the 
creation of the moon \citep{canup2000origin}. 
Our moon is exceptionally large and it is established
that its size helps to stabilize the rotation axis
(the obliquity) of earth \citep{laskar1993stabilization}. 
Seasonal variability (between sumer and winter) would
be otherwise substantially larger. The presence of
the moon is however not a precondition for 
habitability per se. 

\subsection{4.5-4.0\,Ga: The hadean CO$_2$ sequestration
            \label{sec:hadean_CO_2}}

It is not known how wet earth initially was, viz
which percentage of today's water was initially
present and to which extent water was brought
to young earth from further-out solar objects
\citep{drake2002determining}. It is however 
believed that the outgasing of the volatiles from the 
initially hot magma lead to a dense CO$_2$ atmosphere
\citep{sleep2010hadean} (about 100\,bar of CO$_2$, as for 
today's Venus), which, in turn, prevented earth to cool 
below 500$\degree$C after the formation of the moon. 
A liquid ocean of some extent would however been present
despite the elevated surface temperature as a consequence
of the likewise increased atmospheric pressure.

Any planet hoping for an earth-like habitability 
needs to rid itself of its primordial CO$_2$ 
atmosphere. This was achieved on earth by the
carbonation
\begin{equation}
\mathrm{CaSiO}_3\, +\, \mathrm{CO}_2 \ \to\
\mathrm{CaCO}_3\, +\, \mathrm{SiO}_2
\label{eq:Urey}
\end{equation}
of silicates and the subsequent subduction of carbonized 
rocks (the Urey weathering reaction). It is unsure how long 
it actually took, possibly up to the end of the Hadean 
(4\,Ga), for the subducted crust to sequester CO$_2$ to levels 
of only 10-100 times today's levels (about 0.0004\,bar). 

The hadean CO$_2$ sequestration was a massive process.
For a perspective we note that the present day rate of
CO$_2$ sequestration by modern plate tectonics is
of the order of $3.3\times10^{18}$\,mol/Ma \citep{sleep2001carbon}.
With 100\,bar CO$_2$ corresponding to 
$12000\times10^{18}$\,mol this implies, that the
sequestration of the primordial CO$_2$ at present-day 
rates would have taken $(12000/3.3)$\,Ma, viz $3.6$\,Ga.
Substantially faster subduction processes must have
been consequently at work in the early Hadean
\citep{sleep2014terrestrial}.

\begin{table*}[t]
\setlength\arrayrulewidth{2pt} \arrayrulecolor{darkOrange}
\begin{tabular}{rr|lrrrr}
Z   &    & element      &    solar & earth    & crust       & body     \\
\hline
\rowcolor{lightYellow}
1   & H  & Hydrogen     &        [*]  &  0.67  &  2.9   & 62.5   \\
\rowcolor{lightGreen}
2   & He & Helium       &  [$^\sharp$]&          &          &          \\
\rowcolor{lightBrown}
6   & C  & Carbon       &       25.05 &  0.16  &  0.035 & 11.6   \\
\rowcolor{lightYellow}
7   & N  & Nitrogen     &        7.39 &  0.005 &  0.003 &  1.2   \\
\rowcolor{lightGreen}
8   & O  & Oxygen       &       54.70 & 48.3   & 59.9   & 24.1   \\
\rowcolor{lightBrown}
12  & Mg & Magnesium    &        3.59 & 16.5   &  2.0   &  0.007 \\
\rowcolor{lightYellow}
13  & Al & Aluminum    &        0.29 &  1.5   &  6.3   &          \\
\rowcolor{lightGreen}
14  & Si & Silicon      &        3.48 & 15.0   & 20.9   &  0.006 \\
\rowcolor{lightBrown}
15  & P  & Phosphorus   &        0.03 &  0.1   &  0.07  &  0.22  \\
\rowcolor{lightYellow}
16  & S  & Sulfur       &        1.47 &  0.52  &  0.023 &  0.04  \\
\rowcolor{lightGreen}
19  & K  & Potassium    &        0.01 &  0.01  &  1.1   &  0.03  \\
\rowcolor{lightBrown}
20  & Ca & Calcium      &        0.21 &  1.1   &  2.2   &  0.22  \\
\rowcolor{lightYellow}
26  & Fe & Iron         &        2.95 & 14.9   &  2.1   &  0.0007\\
\end{tabular}
\hspace{2ex}
\begin{minipage}{0.45\textwidth}
\caption{
\label{table_elements}
The mol-abundances (relative number of atoms, not weight; in
 percentage) of some selected elements. For the solar system 
(disregarding Hydrogen [*] and Helium [$^\sharp$]) \citep{lodders20094}, the earth 
(all) \citep{mcdonough1995composition}, 
the crust (of the earth) \citep{lide2008crc}, and for the human body 
\citep{lide2008crc}. Note that the curst is weakly reducing in the 
sense that the available oxygen is nearly enough to oxidize all 
other elements via $\mathrm{Si}+\mathrm{O}_2\to \mathrm{SiO}_2$,
$4\mathrm{Al}+3\mathrm{O}_2\to 2\mathrm{Al}_2\mathrm{O}_3$,
etc. 92.1\% and 7.8\% of the overall number of atoms of the
solar system are Hydrogen and Helium atoms respectively.
        }
\end{minipage}
\end{table*}

It is reasonable to expect CO$_2$ sequestration to be
generically vigorous on potentially habitable rocky planets.
\begin{itemize}
\item The essentially unlimited reservoir of silicates rocky 
      planets like the earth dispose of, compare Table~\ref{table_elements},
      allows to sequester basically all CO$_2$ from the 
      atmosphere. The steady-state level of atmospheric CO$_2$ 
      will then result on habitable planets from the balance
      between volcanic outgasing and ongoing sequestration
      (the inorganic carbon cycle) \citep{sleep2001carbon}.
\item The hadean CO$_2$ sequestration occurred at a time when
      earth dissipated its internal heat by bottom-up mantle 
      convection, viz when modern plate tectonics was most probably 
      not yet at work \citep{korenaga2013initiation}. 
      We will come back to this issue in Sect.~\ref{sec:SLHP}.
\end{itemize}

It is not clear whether Venus had ever been able to start the 
process of CO$_2$ sequestration even when starting out
with earth-like conditions. It may have been, that its
initial magma ocean took substantially longer time (100\,Ma 
instead of 1-4\,Ma, as for earth) to solidify \citep{hamano2013emergence}
and that Venus may have had lost most of its primordial water 
by that time through hydrodynamic escape
\citep{hamano2013emergence,leconte2013increased,kasting2015stratospheric}.
The mechanism involves water to rise to the stratosphere and to 
photodissociate via $\mathrm{H}_2\mathrm{O}+\mathrm{light}\to H_2 + O$.
The free hydrogen molecules then escape the pull of gravity, 
being too light for a rocky planet, dragging some of the oxygen 
with it. This process led on Venus to the loss of an ocean 
worth of water, shutting down also the carbonation 
of silicate rocks via the Urey weather reaction (\ref{eq:Urey}),
which needs in turn the formation of carbonic acid 
$\mathrm{CO}_2+\mathrm{H}_2\mathrm{O} \to\mathrm{H}_2\mathrm{CO}_3$ 
and hence the presence of liquid water in an intermediate step.
The concentration of hydrogen bearing molecules like 
$\mathrm{H}_2\mathrm{O}$ and $\mathrm{NH}_4$ is on the other 
side very low in the stratosphere of earth, at least nowadays, 
and such the loss of $\mathrm{H}_2$ \citep{catling2001biogenic}.

\subsection{3.9-3.8\,Ga: The late heavy bombardment
           \label{sec:LTB}}

After the formation of the moon not much happened
apart from the ongoing CO$_2$ sequestration for 
about 600\,Ma. Then, by 3.9-3.8\,Ga, the late
heavy bombardment (LHB) took place in the form of 
a cataclysmic wave of planetesimals (both
asteroids and comets) battering earth together 
with the entire inner solar system \citep{gomes2005origin}.
An event like the LTB would eradicate all higher life 
forms on a planet, if existing, but it would not
sterilize the planet altogether. The LTB may have
been delivering in addition a certain fraction of 
the water present nowadays on earth ($78\times 10^{21}$ 
mol H$_2$O in the ocean alone), without affecting 
otherwise the habitability of the planet.
 
\begin{figure*}[t]
\centering
\includegraphics[width=0.75\textwidth,angle=0]{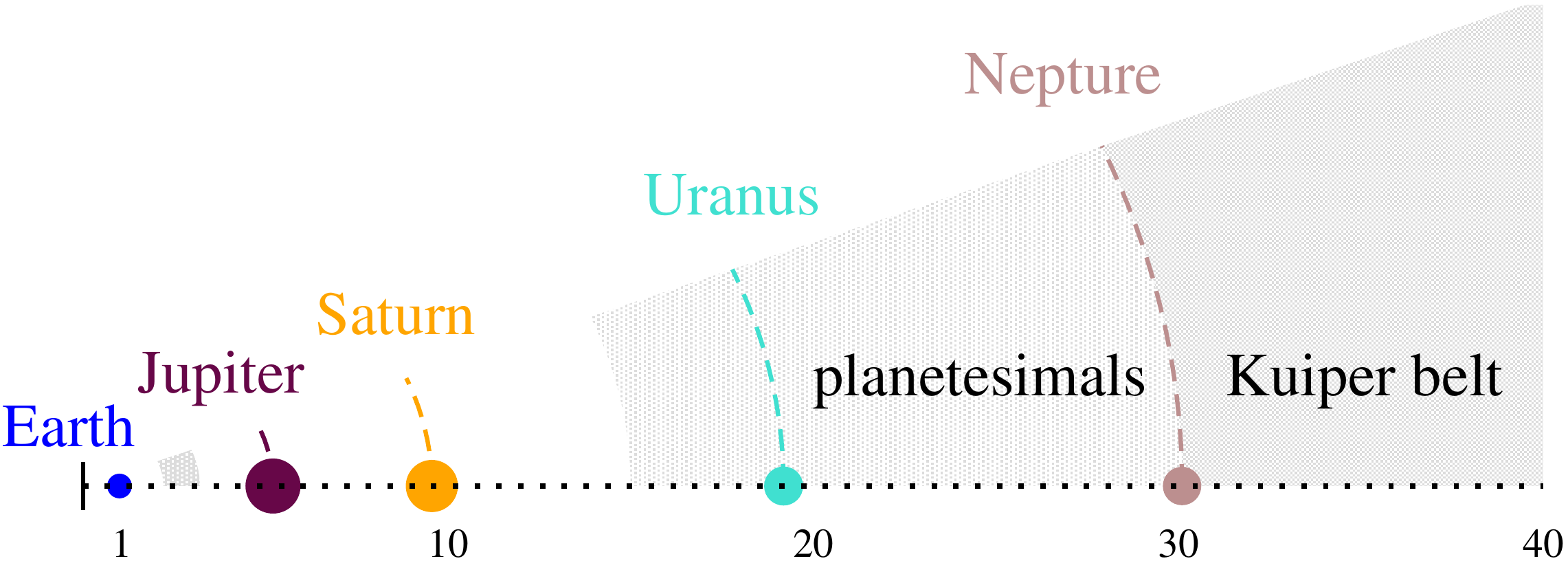}
\caption{An illustration for a possible scenario for the origins
of the great late bombardment. Shown are today's planet positions
(on scale) together with the location of the modern Kuiper belt
of comets and a putative initial belt of unused planetesimals
(a then still present leftover of the solar system formation). 
The mutual interaction with these disks may have caused Saturn 
and Jupiter to migrate until they entered a 2:1 orbital resonance. 
Today's orbital periods of 29.5 and 11.9 years, respectively for 
Saturn and Jupiter, are actually closer to a 3:1 resonance. An 
eccentricity instability would have occurred, roughly 700\,Ma after 
the formation of the solar system, leading to a disturbance of 
both the disk of planetesimal and of the asteroid belt between 
Mars and Jupiter \citep{gomes2005origin}.
The distances are given in astronomical units (earth's distance
from the sun, AU).
        }
\label{fig:solarSystem}
\end{figure*}

The late heavy bombardment did originate, as far as we
know, from an instability of a disc of planetesimals
left over from the formation of the solar system. What is 
interesting is, that such disks must have continued to 
exist long after the formation of the solar system and 
that the instability occurred with a huge delay. A 
possible scenario for this to happen is illustrated 
in Fig.~\ref{fig:solarSystem} (the Nice model 
\citep{gomes2005origin,tsiganis2005origin,bottke2012archaean}). 
It assumes that a 2:1 orbital resonance (in terms of their 
respective orbital periods) did built up due to the migration 
of Jupiter and Saturn, which were in turn caused by the 
interaction with the then still existing outer discs of 
planetesimals. The orbits of both Jupiter and Saturn were 
strongly deformed at resonance, with the consequence that 
the disc of planetesimal was perturbed together with the 
then substantially denser disc of asteroids between Mars 
and Jupiter. Objects were subsequently thrown out of their 
original orbits and sent into the inner solar systems.

\subsection{3.3-2.9\,Ga: The archean genetic expansion
            \label{sect:archean_genetic_expansion}}

Life emerged on earth in a multi-step process which may 
have started quite soon after the late heavy bombardment 
\citep{bada2004life}, possibly also before, having come 
to a completion by around 3.5\,Ga \citep{nisbet2001habitat}. 
Using phylogenomic methods to analyze the evolutionary 
history of 3983 gene families it was found 
\citep{david2011rapid} that the {\it de novo} creation 
of bacterial genes was a concentrated process, the archean 
genetic expansion, starting around 3.3\,Ga and ending 
by 2.9\,Ga with the essential completion of the
bacterial molecular machinery. Gene transfer and 
loss tended to dominate bacterial evolution ever since 
\citep{david2011rapid} and it is not understood why it took
evolution another Ga to develop eukaryotic cells
(see Sect.~\ref{sec:cambrian_explosion}).

\subsection{2.4-2.3\,Ga: The great oxidation event
            \label{sec:GOE}}

Living organisms need an energy source to power
reaction pathways utilizing $\mathrm{CO}_2$ for 
the synthesis of organic molecules (carbon fixation). 
Early in the history of life this energy was provided mostly 
by chemotrophic reaction pathways \citep{wachtershauser2006volcanic},
such as
\begin{equation}
3\mathrm{FeS} + 4\mathrm{H}_2\mathrm{S} + \mathrm{CO}_2 \ \to\  
3\mathrm{FeS}_2 + \mathrm{CH}_3\mathrm{SH} + 2\mathrm{H}_2\mathrm{O}~,
\label{eq:chemotrophy}
\end{equation}
which uses in this case the energy released by the reaction of 
ferrous sulfide ($\mathrm{FeS}$) with hydrogen sulfide 
($\mathrm{H}_2\mathrm{S}$). The reaction products are here 
pyrite ($\mathrm{FeS}_2$), methanethiol ($\mathrm{CH}_3\mathrm{SH}$) 
and water ($\mathrm{H}_2\mathrm{O}$).

The global bioproductivity of chemotrophy is limited by the 
rate by which volcanism replenishes the primary reagents. 
For present day hydrothermal activity levels this implies
that about $(2-20)\times10^{12}$ mol organic C per year 
could be fixated via chemical pathways \citep{des2000did}. Today's 
biosphere produces by contrast around $8700\times10^{12}$ mol 
organic C per year using the light of the sun to power the 
photosynthesis reaction
\begin{equation}
6\mathrm{CO}_2 + 6\mathrm{H}_2\mathrm{O} + \mathrm{light}
\ \to\  \mathrm{C}_6\mathrm{H}_{12}\mathrm{O}_6 + 6\mathrm{O}_2~.
\label{eq:photosynthesis}
\end{equation}
The invention of photosynthesis occurring by the end of the 
archean genetic expansion, around 2.8-2.5\,Ga \citep{des2000did},
possibly also later \citep{rasmussen2008reassessing}, 
did therefore allow the terrestrial biosphere to expand 
dramatically. The balance equation (\ref{eq:photosynthesis}) 
is in addition oxygenic in the sense that free $O_2$ is produced
as a waste besides glucose ($\mathrm{C}_6\mathrm{H}_{12}\mathrm{O}_6$). 

Earth changed in many ways once life started to produce oxygen in
relevant quantities \citep{catling2001biogenic}.
\begin{itemize}
\item The $\mathrm{Fe}^{+2}$ ions present till then in the ocean were 
      rapidly precipitated as banded iron formations \citep{lyons2014rise}.
      Essentially all iron-based biological pathways, like anoxic 
      photosynthesis \citep{canfield2005early} (which does not
      produce oxygen), came hence to an abrupt end, at least on
      a global scale, surviving only in suitable niches.
\item Non-$\mathrm{CO}_2$ greenhouse gases like methane
      ($\mathrm{CH}_4$) were equally washed out from the
      atmosphere. The sky became clear and earth the blue
      planet of today. Global temperatures dropped consequently. 
\end{itemize}

Taking a closer look at the elements making up our home
planet, as listed in Table~\ref{table_elements}, one
notices that the crust is weakly reducing in the sense 
that the oxidation of the other elements, 
$$
\mathrm{Si}+\mathrm{O}_2\to \mathrm{SiO}_2,\quad\quad
4\mathrm{Al}+3\mathrm{O}_2\to 2\mathrm{Al}_2\mathrm{O}_3,
\quad\quad\mathrm{etc.}~,
$$ 
would need somewhat more than the actually available amount
of oxygen (earth as a whole would be on the other hand strongly 
reducing). For oxygen to accumulate in the atmosphere two 
things must consequently happen.

\begin{figure}[t]
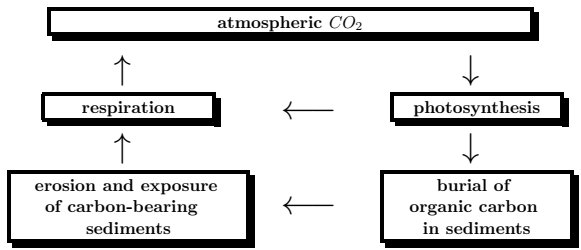

\centering
\setlength\arrayrulewidth{0pt} \arrayrulecolor{darkOrange}

\scalebox{0.65}{
\begin{tabular}{ccc}
\multicolumn{3}{c}{
\myShadowBox{\parbox{9cm}{\centering\bf atmospheric $CO_2$}\vspace{0.5ex}}
                   } \\[1ex]
\scalebox{2}{$\uparrow$} & & \scalebox{2}{$\downarrow$} \\[1ex]
\myShadowBox{\parbox{2.6cm}{\centering\bf respiration}\vspace{0.5ex}} & 
\raisebox{1.0ex}{\scalebox{2}{$\longleftarrow$}} & 
\myShadowBox{\parbox{2.6cm}{\centering\bf photosynthesis}\vspace{0.5ex}} \\
\scalebox{2}{$\uparrow$} & & \scalebox{2}{$\downarrow$} \\[1ex]
\myShadowBox{\parbox{4cm}{\centering\bf erosion and exposure\newline 
                          of carbon-bearing \newline sediments}\vspace{1ex}}
 & \raisebox{4.5ex}{\scalebox{2}{$\longleftarrow$}} & 
\myShadowBox{\parbox{3.0cm}{\centering\bf \ \ burial of\newline
                         organic carbon \newline in sediments}\vspace{1ex}}
\end{tabular}
              } 
\caption{Sketch of the organic carbon cycle
\citep{jacob1999introduction,sleep2001carbon}.
The overall amount of sedimentated organic carbon 
has been estimated to be of the order of 
$1250 \times 10^{18}\,\mathrm{mol}$. This implies,
with about $37 \times 10^{18}\,\mathrm{mol}$ $O_2$ 
present nowadays in the air, that earth's biosphere 
has produced at least $1250/37\approx34$ times more 
oxygen over the last four billion years than the one 
remaining in today's atmosphere.\newline
The present-day fluxes are: 
$8.7 \times 10^{15}\,\mathrm{mol}$ biological C per 
year global NPP (net primary production) \citep{field1998primary}, 
with roughly equal contributions from land and 
oceans, and $11\times10^{18}\,\mathrm{mol}$ 
biological C/Ma buried in sediments \citep{sleep2001carbon}.
        }
\label{fig_organic_carbon_cycle}
\end{figure}

\begin{itemize}
\item Part of the carbon fixated via oxygenic photosynthesis
      must be removed from the surface via sedimentation in
      a process denoted the organic carbon cycle (see 
      Fig.~\ref{fig_organic_carbon_cycle}).
\item All reducing elements (like iron) present on the surface must
      first be oxidized. 
\end{itemize}

It is notoriously difficult to determine how, when and why
oxygen did eventually accumulate in the atmosphere 
\citep{catling2005earth,lyons2014rise}. It is however clear,
that oxygen appeared at appreciable levels (of the order
of a few percent of today's levels) in the atmosphere
during the great oxidation event around 2.4-2.3\,Ga.
Oxygen levels didn't though remain stable afterwards,
possibly plunging again for a billion years or more,
with modern levels of $\mathrm{O}_2$ being reached only
at the end of Proterozoic, viz shortly before the 
cambrian explosion. This second rise of $\mathrm{O}_2$ levels
was the prerequisite for the subsequent development of 
animals and hence all important from a human perspective. 
Its ultimate causes are however not yet understood 
\citep{och2012neoproterozoic}.

\subsection{540-520\,Ma: The cambrian explosion
            \label{sec:cambrian_explosion}}

Though it is difficult, it is not impossible for bacteria 
to develop into at least primitive multicellular organisms
\citep{shapiro1997bacteria}. All higher plants and 
animals are built however out of eukaryotic cells,
which dispose of a much higher complexity of internal
organization. Eukaryotes evolved out of prokaryotes
(bacteria and archaea) in a multi-stage process
ending, as determined e.g.\ by a multigene molecular clock 
analysis \citep{parfrey2011estimating}, by around 
1.9-1.7\,Ga. At that point, the last common ancestor 
of all eukaryotes drifted through the waters of the 
Proterozoic \citep{knoll2006eukaryotic}.

Very little is known when the next important step
in the evolution of life, sexual reproduction, did 
take place, \citep{goodenough2014origins}. It could 
even be the case that no additional step was needed, 
viz that sexual reproduction is inherent to eukaryotic 
life per se \citep{speijer2015sex} and that the last common 
eukaryotic ancestor was already sexual. The advantages 
of sexual reproduction for multicellular organisms are 
in any case undisputed.

The pace of evolution did not pick up directly with 
the invention of eukaryotic cells. Only in the second
part of the following about 1.3\,Ga, the `boring billion' 
\citep{roberts2013boring}, animal phyla started to
diverge measurably \citep{blair2005molecular}. It is 
unknown \citep{zhang2014triggers} which of the
geo-biological events accompanying the demise of this 
prolonged period of stasis where the actual drivers both 
for ending the boring billion and for initiating the 
following unprecedented divergence of life known
as the `cambrian explosion'
\citep{morris2000cambrian,marshall2006explaining}.

Multicellular organisms developed not in a singular 
event but at least on 25 distinct occasions 
\citep{grosberg2007evolution},
with the first major multicellular biota emerging being 
the ediacaran fauna (590-540\,Ma). Of all animals crawling
on the earth however only the Ecdysozoa (like centipedes) 
have ediacaran ancestors \citep{rota2013molecular}. All other
animal phyla appeared in contrast during the following 
cambrian explosion, with the initial burst (540-520\,Ma)
lasting only 20\,Ma. Soon after land was colonized by 
the ancestors of our modern plants and animals (510-470\,Ma) 
\citep{rota2013molecular}.

\subsection{20\,Ma: C4 photosynthesis 
            \label{sec:C4}}

Earth started, as discussed in Sec.~\ref{sec:hadean_CO_2},
with something like 100\,bar of $\mathrm{CO}_2$, which were
then rapidly sequestrated. Volcanic outgasing and limited
sedimentation rates kept atmospheric $\mathrm{CO}_2$ 
concentrations afterwards at levels which where still 
relative high in comparison to today's value \citep{kaufman2003high}. 
By 2.2\,Ga atmospheric $\mathrm{CO}_2$ was, for a reference, 
about $23\times\mathrm{PAL}$ (present day preindustrial
atmospheric levels: 280\,ppm) \citep{sheldon2006precambrian}. 
That changed however when global bioproductivity increased
after life colonized land in the aftermath of the cambrian 
explosion \citep{igamberdiev2006land}, both because
of the additional carbon fixation by the land plants
and because the roots of the plants intensified, in 
addition, the weathering of rocks and hence the amount
of biologically available mineral nutrients (like phosphorus).
The resulting decline of atmospheric $\mathrm{CO}_2$
led eventually to such low levels of $\mathrm{CO}_2$ (about 
present-day PAL) \citep{sage2012photorespiration}, that 
life had to readjust.

The reason is that C3 photosynthesis, the dominant pathway 
for oxygenic photosynthesis for plants, becomes at first 
linearly less effective with declining $\mathrm{CO}_2$ 
concentration \citep{collatz1977influence}, stopping in the 
end altogether once the $\mathrm{CO}_2$ level falls below 
a certain threshold (which depends in turn on other parameters
like humidity, oxygen level and the like). The recent, 
human-induced raise of atmospheric $\mathrm{CO}_2$ has led 
conversely to an ongoing greening of earth
\citep{sheldon2006precambrian}. By 20\,Ma, with precursors 
starting around 30\,Ma, C4 photosynthesis was developed
as a new pathway for photosynthesis by at least 66
different terrestrial plants \citep{sage2012photorespiration}. 
The efficiency of C4 photosynthesis does not depend on 
$\mathrm{CO}_2$ partial pressures, in contrast to C3 
photosynthesis, and it is therefore believed that its evolution 
constitutes a response of the biosphere to a chronic shortage 
of atmospheric carbon dioxide \citep{sage2012photorespiration}.

\begin{figure}[t]
\includegraphics[width=0.45\textwidth,angle=0]{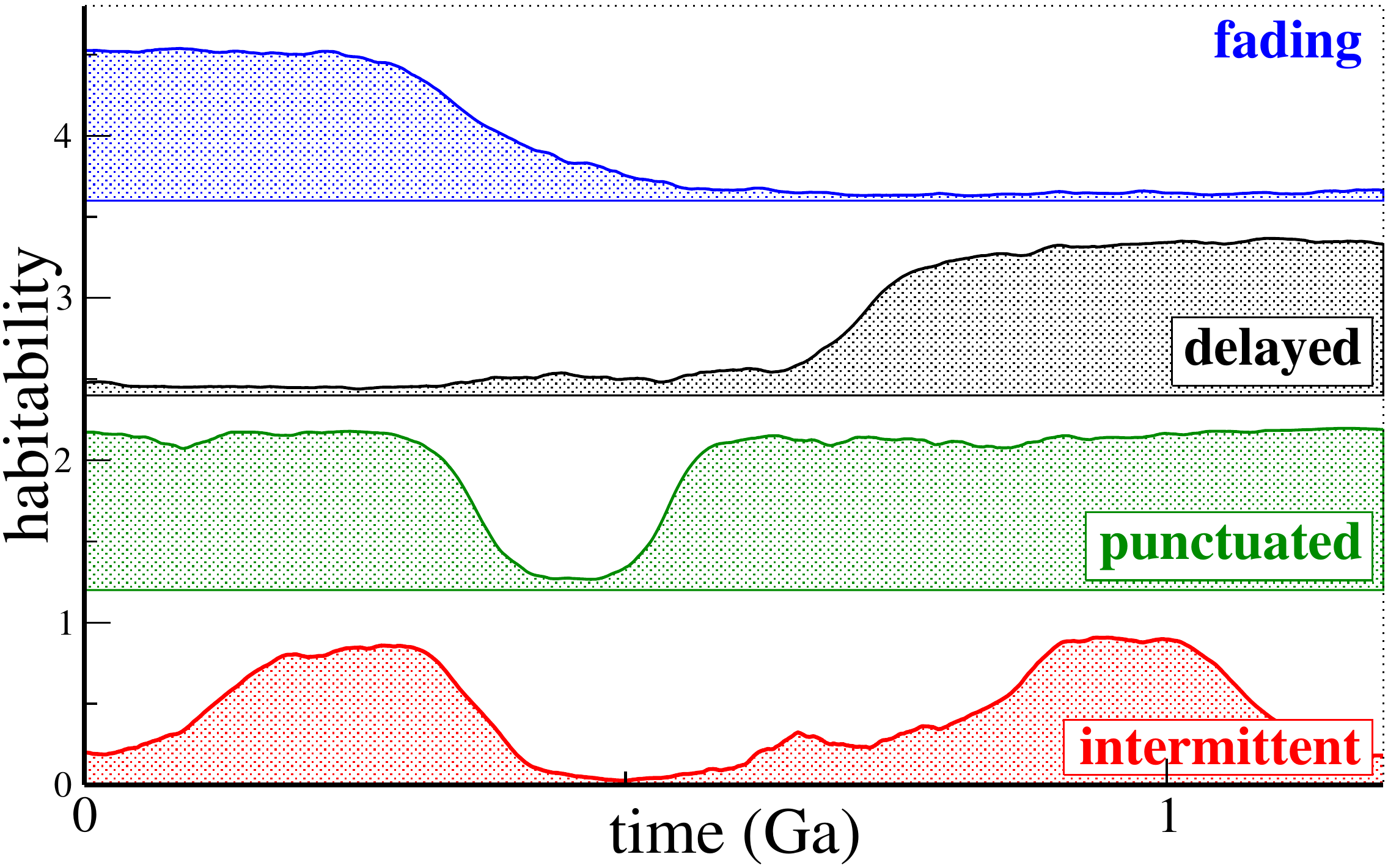}
\caption{Schematic illustration of the four possible classes
         of transient habitability. The timescale may range
         from (1-2)\,Ga for a FGK host star or up to
         10\,Ga for a planet orbiting a M dwarf.
        }
\label{fig:habitability}
\end{figure}

Today about 23\% of the terrestrial NPP (net primary production
of organic carbon) is due to C4 plants. During the last
0.4\,Ma, when the $\mathrm{CO}_2$ level oscillated between
180\,ppm (during periods of extended glaciation) and 
280\,ppm (during interglacials) \citep{petit1999climate}, 
an additional expansion of C4 vegetation could be observed
every time $\mathrm{CO}_2$ levels dropped to
200\,ppm or below \citep{pinto2014photosynthesis}. One can 
regard the emergence of C4 photosynthesis hence as a turning 
point in the history of our planet, in the sense that earth's 
biosphere will be entrenched, from now on till the end, in a battle 
over the ever declining amounts of recycled $\mathrm{CO}_2$ pumped 
out by earth's progressively receding geothermal activity
\citep{o2013swansong}.
The total amount of $\mathrm{CO}_2$ remaining in the atmosphere 
today, $0.05\times10^{18}$ mol at 280\,ppm, is actually so small,
that an individual $\mathrm{CO}_2$ molecule will remain in the
air for only 5-10 years \citep{jacob1999introduction}. Ever hungry
plants are waiting \citep{zhu2016greening}. The residence times 
for $\mathrm{O}_2$ and $\mathrm{N}_2$ molecules are, on the other 
side, 3\,Ma and 13\,Ma respectively.

\begin{figure}[t]
\centering
\includegraphics[width=0.45\textwidth,angle=0]{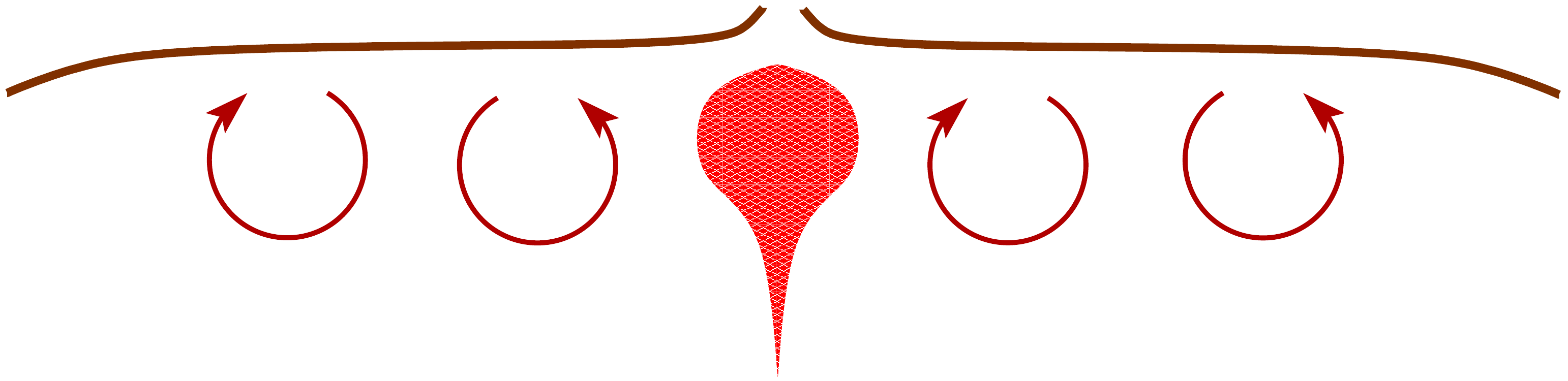}\\[3ex]
\includegraphics[width=0.45\textwidth,angle=0]{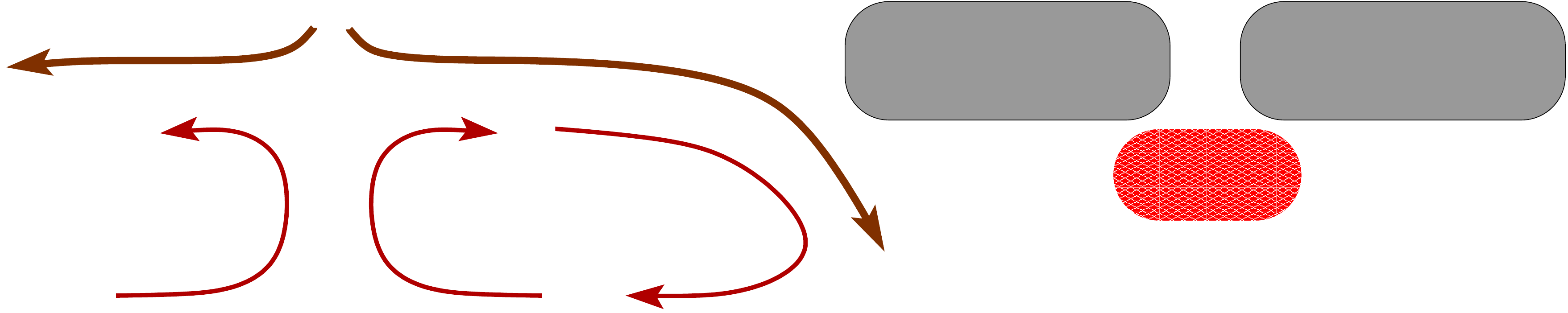}
\caption{{\em Top}: Illustration of stagnant-lid tectonics, for which
          lithosphere acts as a lid for the mantle dynamics. Updwelling 
          plumes will however erupt through the lid time and again.\newline
          {\em Bottom}: Illustration of plate tectonics. The ocean floor 
          is subducted below the continental plates (grey), which
          tend in turn to act as heat traps. Mantle may overheat
          especially below a supercontinent, rise and erupt
          in the form of extended basaltic flooding.
        }
\label{fig:tectonics}
\end{figure}

\subsection{Earth's lost billions}

Evolution is seldomly dependent on singular events.
This holds also for major evolutionary steps like
the invention of multicellularity and C4 photosynthesis,
which have been developed independently, as discussed in 
Sec.~\ref{sec:C4} and \ref{sec:cambrian_explosion}, 
on at least 25 and 66 occasions respectively. Why did earth
then take about one Ga to develop full fledged
bacteria, and another one for the eukaryotic cell?
It could not have been for the lack of living
organisms.

It has been estimated in this context \citep{kallmeyer2012global}
that about $(9-32)\times10^{29}$ bacteria dwell
on earth, nowadays, making up in turn a few percent of 
the overall biomass. Given or taken a few orders
of magnitude, we may assume that a similar number
of microbes populated earth since the inception
of life, with the population density being somewhat 
smaller before the invention of oxygenic photosynthesis 
in the late Archean (compare Sect.~\ref{sec:GOE}).
$10^{29}$ organism with a life cycle of hours to
days harbor an enormous evolutionary potential
and a better understanding of the mechanism
tapping into this evolutionary potential would 
hence help to clarify greatly our perspective of 
finding complex life elsewhere in the universe 
\citep{ward2000rare}. The alternative view taken here 
is that habitable planets would have a much higher 
chance to bear complex life if they would be given 
a speedup by several Ga.

\section{Habitability that waxes and wanes\label{sec:SLHP}}

Transient habitability can be classified into four 
fundamental types (compare Fig.~\ref{fig:habitability}):

\begin{itemize}
\setlength{\itemsep}{-2pt}
\item[--] fading habitability, when initially clement conditions become
    progressively adverse,
\item[--] delayed habitability, if the planet becomes habitable only late 
    in the lifetime of the host star,
\item[--] punctuated habitability, if the habitability is interseeded
    by relatively short periods of inhabitable conditions and
\item[--] intermittent habitability, whenever clement conditions stabilize
          only intermittently.
\end{itemize}

Punctuated habitability is in this context the type 
of habitability taking a prominent place in doomsday 
scenaria. There is however a difference between 
catastrophic extinction and the temporarily termination 
of habitability per se. For a perspective we recall that 
the permian mass extinction at 252\,Ma \citep{benton2003life}, 
the biggest extinction event ever occurring on earth, erased 
only 80-96\% of the marine and about 70\% of the terrestrial 
vertebrate species. Biodiversity had no problems to recover 
in the aftermath within 8-9\,Ma together with overall trophic 
levels \citep{chen2012timing}.

It is actually surprisingly hard to construct a scenario in 
which a singular event wipes multicellular life completely off 
the surface of a planet. The 1-1000\,sec gamma-ray bursts 
emitted (possibly as frequent as once every 500\,Ga in
our galactic habitat \citep{piran2014possible}) during the
collapse of a nearby massive star, have been discussed 
extensively in this context together with other conceivable
high-energy astrophysical events \citep{horvath2012effects}.
Such a burst would indeed sterilize half the planet, viz
the exposed face, leading in addition to a series of
events which would deplete the ozone layer for a few
months by an average of 40\% \citep{thomas2005gamma}. For 
comparison we note that the areas affected by the present-day 
antarctic ozone hole, on earth, which itself has had depletion 
levels of up to 50\%, have actually seen only a reduction of 
terrestrial plant productivity by less than 6\% 
\citep{ballare2011effects}. 

\subsection{Stagnant-lid planets}

Mantle and crustal reorganization processes are driven 
by the need to dissipate the internal heat. They often 
settle into steady-state convection patterns
\citep{ernst2009archean}, with the two most important 
types being stagnant-lid and plate tectonics 
(compare Fig.~\ref{fig:tectonics}).

Plate tectonics ensures on earth the continuous 
recycling of carbon, which would be otherwise lost
as a consequence of the inevitable sedimentation of
biomass and due to the ongoing carbonization of the 
oceanic lithosphere. Carbon is released back to the 
atmosphere from the subducted ocean floor in part 
directly and in part indirectly, through further 
mantel convection, by arc volcanos and by the 
mid-oceanic ridges respectively \citep{sleep2014terrestrial}. 
The oceanic lithosphere is such fully renewed
within 170\,Ma, recycling on the way about
$3.3\times10^{18}$\,mol carbon per Ma
\citep{sleep2001carbon}. The sequestration of a 
100\,bar worth of carbon dioxide through the burial 
of carbonated crust, compare Sect.~\ref{sec:hadean_CO_2}, 
was however achieved in the Hadean without the help 
of plate tectonics, which was then not yet operative 
\citep{harrison2009hadean}.

Plate tectonics is the dominant but not the only type of 
tectonic activity present on our home planet. Other
known processes are mid-plate volcanism, such as the one
causing the Hawaiian and Yellowstone hotspots 
\citep{fouch2012yellowstone}, and the subcontinental 
overheating of magma (or the updwelling of mantle plumes 
\citep{santosh2014cambrian}), which is thought to be 
causal for the widespread basaltic flooding occurring 
in conjunction with the breakup 
of supercontinents \citep{coltice2007global}.

Carbon recycling will be in contrast a much more
dramatic process on planets with stagnant-lid 
tectonics, which do not dispose of a primary mechanism 
for the continuous recycling of $\mathrm{CO}_2$ 
\citep{lammer2009makes}, but most probably of a 
discontinuous carbon cycle in the form of episodic 
basaltic overturns. The resulting fluctuations of 
atmospheric CO$_2$ levels will however not forestall
habitability per se \citep{petit1999climate}, at least
as long as no runaway instability is induced. It has been 
suggested in this context, that stagnant-lid planets may 
have a more vigorous mantle dynamics than planets with 
plate tectonics \citep{kite2009geodynamics} and
that their volcanic activity may abate consequently
somewhat faster. Stagnant-lid planets can therefore 
be expected to to support clement conditions for 
extended but otherwise limited periods and to be 
hence prime candidates for the Genesis mission.

Stagnant-lid planets may of course also be utterly
inhabitable whenever other factors won't allow it. 
The classical example is Venus, where the stagnant 
curst is punctuated continuously by updwelling plumes 
\citep{phillips1998geological}, in part on a local 
and in part on a global scale \citep{smrekar2010recent}.
It is not known whether plate tectonics would have set
in, eventually, in the case that Venus would not have
lost its ocean.

\begin{figure}[t]
\includegraphics[width=0.45\textwidth,angle=0]{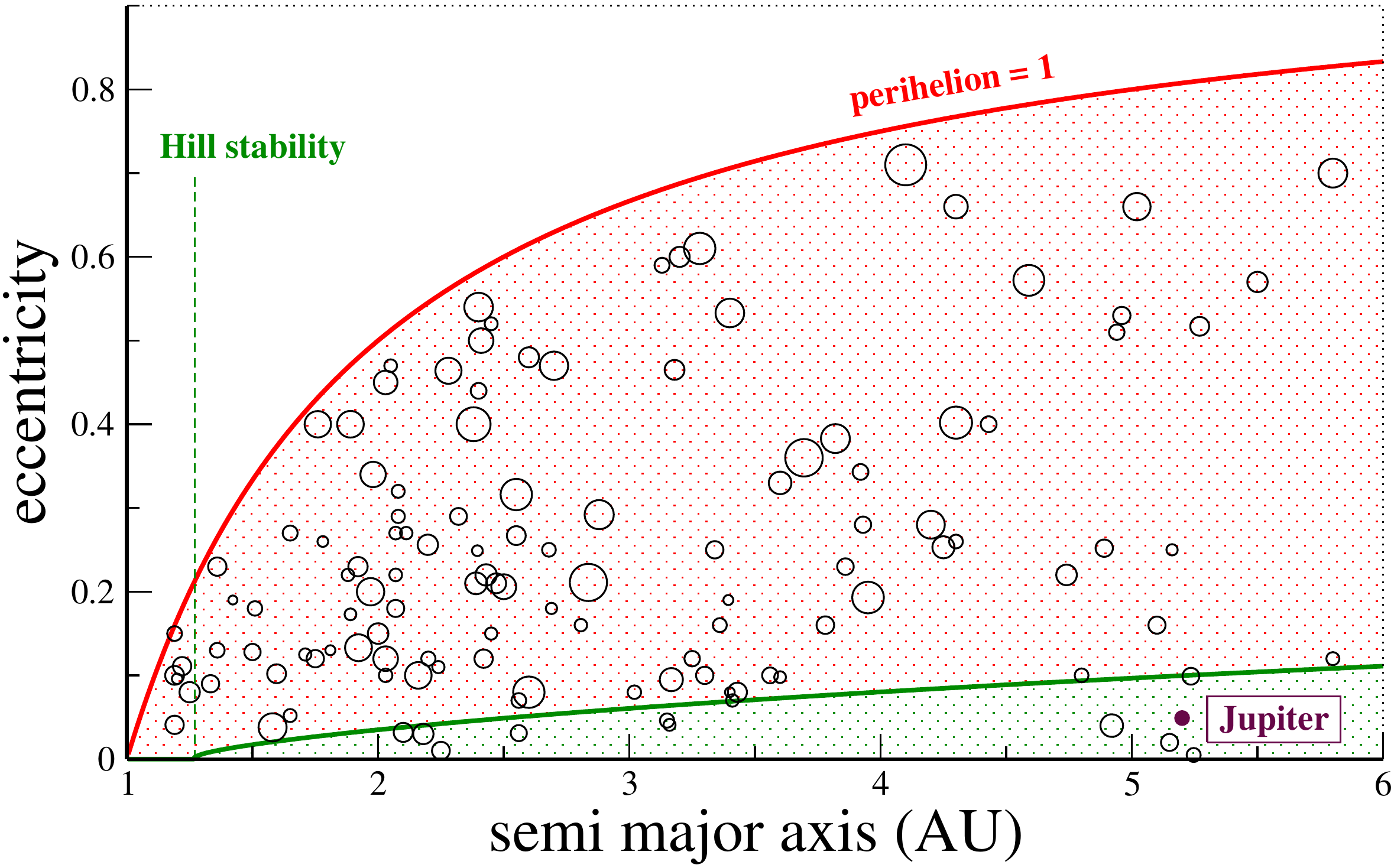}
\caption{The region of Hill stability (below the solid 
green line), as defined by Eq.~(\ref{eq:Hill_stability},
for an earth-like planet ($\mu_2=300\mu_1$) against the 
perturbation by a Jupiter-size planet ($\mu_2=10^{-3}$) with 
eccentricity $e_2$. For large eccentricities the perihelion 
of the Jupiter-like planet would cross the orbit of the 
earth-like planet (solid red line). The open circles are 
129 exoplanet around G and F stars listed in the extrasolar 
planets encyclopaedia \citep{ExtrasolarPlanetsEncyclopaedia},
with the sizes the circles indicating the respective masses. 
The filled maroon dot is the Jupiter of the solar system.
        }
\label{fig:planetsDistanceEccentricity}
\end{figure}

\subsection{Hill instability
            \label{sect:hill_instability}}

A pair of planets is said to be Hill unstable
when their orbits eventually cross due to
their mutual gravitational interaction. The 
resulting massive orbital deformation (if not
a direct collision) would terminate habitability
for any planet located initially in a habitable zone.
For an illustration of this process we consider 
a planetary system with two planets having masses 
$\mu_i$ ($i=1,2$, relative to the mass of the host star) 
and eccentricities $e_i$ (a measure of how elliptic
an orbit is). The Hill stability line
is then determined by \citep{gladman1993dynamics}
\begin{equation}
\left(\mu_1+\frac{\mu_2}{\delta^2}\right)
(\mu_1\gamma_1+\mu_2\gamma_2\delta)^2 \ >\
\alpha^3 + 3^{4/3}\mu_1\mu_2\,\alpha^{5/3}~,
\label{eq:Hill_stability}
\end{equation}
where $\alpha=\mu_1+\mu_2$ is the total relative
mass and $\gamma_i=\sqrt{1-e_i^2}$. The distances
of the two planets to the host stars are taken to
be unity and $1+\Delta$ respectively, with 
$\delta=\sqrt{1+\Delta}$.

Solving Eq.~(\ref{eq:Hill_stability}) for $e_2$ we have 
plotted in Fig.~\ref{fig:planetsDistanceEccentricity}
the stability region for an earth- and Jupiter-like 
planetary system with $\mu_2=300\mu_1$ and $\mu_2=1/1000$.
The influence of $e_1$ is in this case so small that
one can set $e_1\to0$. Overlayed are the parameters of
129 known exoplanets around G and F stars whose 
orbits would not cross the orbit of a putative rocky
planet located at $1\,\mbox{AU}$. Note, that the Hill
instability line shown in Fig.~\ref{fig:planetsDistanceEccentricity} 
has been evaluated only for an outer planet of 
Jupiter mass. It moves up/down for outer planets 
with smaller/larger masses.

Most exoplanetary systems detectable to date are 
dominated by super-Jupiter gas giants. It is hence not
suprising that configurations allowing for Hill stable 
habitable planets are rare.
A certain time is however needed, the orbital lifetime, 
before the instability actually happens. This lifetime 
ranges from typically a few $10^4$ years, deep in the 
unstable region, to up to few $10^9$ years close to the 
Hill stability line 
\citep{rivera2007stability,veras2013simulations}. 
Fig.~\ref{fig:planetsDistanceEccentricity} hence
suggests, that Genesis candidate planets may be 
found in systems with weak Hill instabilities.

Above considerations concerned exoplanets for which 
habitability is eventually terminated. The opposite
may also happen, especially when two or more outer
gas giants scatter dynamically \citep{veras2006predictions}.
The resulting orbital deformations have been shown to
move rocky planets closer to the sun \citep{veras2006predictions},
viz possibly from outside to inside the habitable zone
(in the case that the initial distance was too large). 
A potentially large number of earth-size exoplanets may 
therefore be newcomers to their habitable zone, yet
barren and hence candidates for a Genesis mission.

\subsection{Lagrange instability
            \label{sect:lagrange_instability}}

Genesis candidate planets may also be found in Hill 
stable, but Lagrange unstable planetary systems. A 
massive orbital deformation ocurs also in this
case, for one of the involved planets, this time
however one which leads either to a collision 
with the central star or to an escape from the 
planetary system. Estimating the Lagrange lifetimes 
for a specific exoplanetary system is a demanding 
task \citep{veras2016full} and beyond the scope 
of the present study. We restrict ourselves therefore
to a first assessment whether very long Lagrange 
lifetimes may potentially exist.

For this purpose we consider the minimal time 
$T_L$ for an Lagrange instability to occur.
$T_L$ has been estimated \citep{veras2013simple} 
to scale roughly as
\begin{equation}
\log_{10}\left(\frac{T_L}{T_1}\right) \ \sim\ 5.2\left(
\frac{\mu}{\mu_J}\right)^{-0.18}
\label{eq:langrangeMinimialLifetime}
\end{equation}
for systems composed of two planets with an equal 
relative mass $\mu$ and eccentricities below 
$\sim0.3$. $T_1$ is here the orbital period
of inner planet and $\mu_J$ the relative mass 
of the solar-system Jupiter. This scaling has been 
derived \citep{veras2013simple} for a configuration 
where the relative orbital distance of the two giants 
is confined to within $[1+0.3(1+e_1)(1+e_2)]$ Hill 
limits. The two gas giants are hence assumed to be
in a Hill stable configuration close to the stability threshold.

In a Hill stable three-planet system, with an inner 
rocky planet in the habitable zone and two outer 
gas giants, a Lagrange instability of the inner
gas giant occurring due to the mutual interaction 
between the two gas giants may also throw the 
rocky planet out of its orbit and consequently 
also out of the habitable zone. We have hence 
evaluated 
Eq.~(\ref{eq:langrangeMinimialLifetime})
for all the 91 exoplanets shown in 
Fig.~\ref{fig:planetsDistanceEccentricity}
having an eccentricity less than $0.3$. That
is, we have added by hand to each of these exoplanets
a putative equal mass companion located further out.

\begin{figure}[t]
\hspace{2ex}
\includegraphics[width=0.45\textwidth,angle=0]{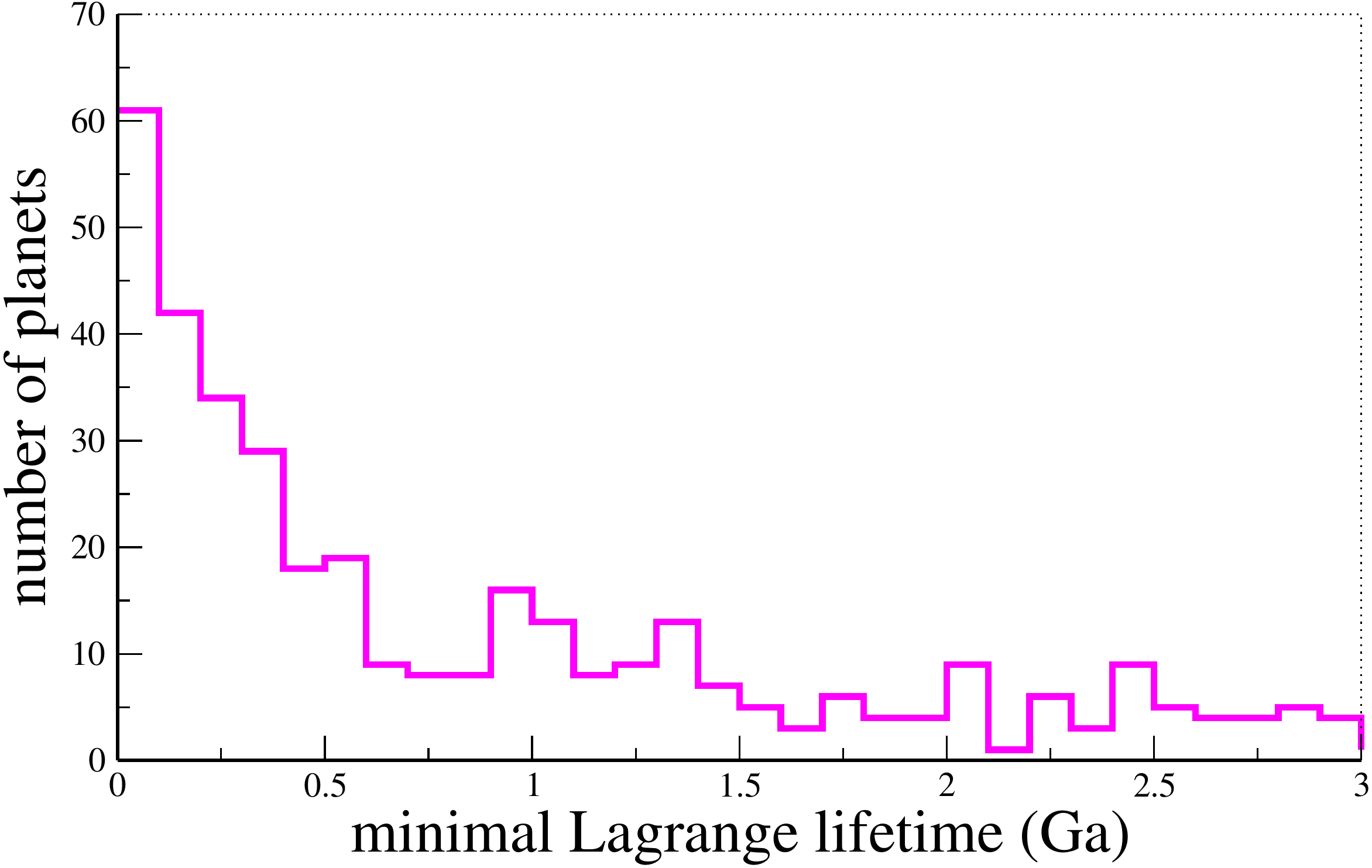}
\caption{The distribution of the minimal Lagrange lifetimes 
(in Ga; bin-size: 0.1\,Ga), as given by
Eq.~(\ref{eq:langrangeMinimialLifetime}), for 91 
exoplanets around G and F stars, with eccentricities 
below $0.3$ and with orbits outside $1\,\mathrm{AU}$,
listed in the extrasolar planets encyclopaedia 
\citep{ExtrasolarPlanetsEncyclopaedia}. The results are 
for putative exoplanetary systems, when a not too far 
away further-out Hill stable equal mass companion would 
exist in addtion to each of the actually observed exoplanet.
        }
\label{fig:minimalLagrangeLifetime}
\end{figure}

The distribution of Lagrange escape times resulting
from this Gedanken-experiment is shown in
Fig.~\ref{fig:minimalLagrangeLifetime}.
Small minimal Lagrange lifetimes clearly rule out
the actual existence of a further-out (but close-by)
equal mass companion in the respective exoplanetary 
system. Fig.~\ref{fig:minimalLagrangeLifetime} shows 
on the other side that long Lagrange escape times
would be present for suitable configurations of
gas giants. Transient habitability as resulting from 
long-term Lagrange- or other orbital instabilities,
like resonances (see Sect.~\ref{sec:LTB}), of the 
host exoplanetary system may hence be common. A 
detailed analysis of the orbital past and future of 
extrasolar systems constitutes therefore an important 
screening tool for prospective Genesis missions.

\section{The Genesis mission}

A very large number of habitable planetes may potentially 
exist in our galaxy. Current estimates range from about 
0.06 per sun-like star
\citep{petigura2013prevalence,silburt2015statistical},
to 0.12-0.24 habitable planets per M-dwarf
\citep{dressing2015occurrence}. The problem is however
the distance. There are only 9 extrasolar systems
(containing 14 stars) within 10 light years (lyr)
and about $14\times10^{3}$ stars within 100\,lyr
(as extrapolated from the Gliese \citep{gliese1991preliminary}
and HIPPARCOS \citep{perryman1997hipparcos} catalogue entries).
Among these we may expect up to a few hundred
potentially habitable planets and possibly up to a
few dozen Genesis candidate planets.

\subsection{Genesis vs.\ interstellar exploration}

Interstellar exploration via robotic or manned crafts
can be regarded as a long-term research investment
aiming to increase our scientific knowledge regarding 
the geophysics, the habitability and the astrobiology 
of extraterrestrial planetary systems 
\citep{crawford2009astronomical}. 
An explorative mission would therefore be only realizable 
if the projected mission duration would respect the 
maximal planning horizon of the funding institution, 
which will correlate in turn with typical human life 
expectancies, say a hundred years \citep{projectDragonfly}. 
The Genesis project is however not for human benefit and 
it is consequently irrelevant how long it would take for a
Genesis craft to arrive to the target. A few millenia 
more or less would be negligible on evolutionary time scales.
\begin{itemize}
\item[--] The absence of strict travel time requirements allows
          in first place for reduced cruising velocities. A 
          Genesis probe would therefore need only comparatively
          modest financial and technical resources. 
\item[--] The irrelvance of cruising times also allows to
          consider time-consuming deceleration options, e.g.\ 
          via magnetic and/or electric drag 
          \citep{zubrin1991magnetic,perakis2016combining}.
\item[--] Overall travelling times are however restricted 
          on the technical side by the lifetime of the 
          constituent components.
\end{itemize}
A Genesis probe targeting candidate planets within a 
radius of 100\,lyr may hence take a minimum of $10^3-10^4$ 
years to arrive. The Voyager spacecrafts need in
comparison about $19\times10^3$ years to travel
one light year. A close-up examinantion of the target
planet, the first thing to be done, would then be
followed by the decisive step, the autonomous decision 
whether to start the seeding sequence. The Genesis 
process would not be started if higher life forms would 
be detected from orbit.

The alternative to an in-situ decision taking, waiting 
for a response back from earth, would rely on the other 
hand on the long-term functioning of a trans-generational 
contract. Such a long-term conditioning would however
be a questionable perspective when considering humanity's 
history of political and social turmoil. In this light
it is actually an appealing aspect of the Genesis concept 
that the craft can be designed as a one-shot launch-and-forget 
project. The backtransmission of the in situ collected data
(to anybody still listening) should be considered that 
notwithstanding to consitute an integral part of the mission 
layout.

\subsection{Biosphere incompatibilities
            \label{sect:biosphere_incompatibilities}}

Let us digress for a moment and ask the question:
What may happen, if humanity's dream of a spaceship 
load of human settlers setting foot on a second 
earth would come true? In this case we would bring 
terrestrial life, microbes included, to a planet 
with a biosphere as rich as the one of our home
planet. Both the alien biosphere and the invading 
fragment of the terrestrial biosphere would 
interpenetrate each other and humantity would have
started a non-reversible experiment for which the
outcome will most probably be determined by how 
universal the immune system of the respective 
multicellular organisms are.

The reason is that all multicellular organisms, plants 
and animals alike \citep{rodriguez2012immune}, are
vitally dependent on a performing immune system for 
their defence agains pathogenic microbes. Key to the 
functioning of an immune reaction is the recognition 
of `non-self', which is achieved in turn by the ability 
of the immune systems, at least on earth, to recognize 
certain products of microbial metabolism that are unique 
to microbiota \citep{medzhitov2002decoding}. 
How likely is it then, that `non-self' recognition will 
work also for alien microbes? 

Here we presume, that general evolutionary principles hold. 
Namely, that biological defense mechanisms evolve only when 
the threat is actually present and not just a theoretical 
possibility. Under this assumption the outlook for two 
clashing complex biospheres becomes quite dire.
\begin{itemize}
\item[--] In the best case scenario the microbes of one
  of the biospheres will eat at first through the higher 
  multicellular organism of the other biosphere. Primitive
  multicellular organism may however survive the onslaught
  through a strategy involving rapid reproduction and adaption. 
  The overall extinction rates could then be kept, together
  with the respective recovery times, 1-10\, Ma 
  \citep{chen2012timing}, to levels comparable to that 
  of terrestrial mass extinction events.
\item[--] In the worst case scenario more or less all
  multicellular organism of the planet targeted for human 
  settlement would be eradicated. The host planet would then 
  be reduced to a microbial slush in a pre-cambrian state, 
  with considerably prolonged recovery times. The
  leftovers of the terrestrial and the indigenous
  biospheres may coexist in the end in terms of 
  `shadow biospheres' \citep{davies2009signatures}.
\end{itemize}
It is then clear, that exoplanets harboring multicellular
life should be off-limit for Genesis missions. Expanding
civilizations may find it equally unattractive to settle 
other life-bearing planets in the context of the Fermi 
paradox \citep{webb2015if,gros2005expanding}. Biosphere
incompatiblities may be generic.

\subsection{Ethics and planetary protection}

Exoplanets already harboring a biosphere with
multicellular lifeforms, being them primitive
or advanced, would not be targeted by Genesis
probes. The objective of the Genesis mission is
after all to give life the chance to prosper 
in places where it has not yet a foothold, and 
not to invade and possibly to destroy
existing biospheres (see 
Sect.~\ref{sect:biosphere_incompatibilities}).
Regarding the ethics
of such an endeavor one may ask whether it
is legitimate to bring life to a planet which
will cease anyhow to be habitable in the foreseeable
future. Here we take the stance that death
is no less part of the life cycle than birth,
which is equivalent to saying that the value 
of being alive is not grounded in the avoidance 
of an inescapable death. We do acknowledge,
however, that this is a viewpoint that will not be 
universally shared (somewhat related issues have
been raised in the context of unlimited human life 
extensions
\citep{pecujlija2013eternal,gyngell2015ethics}).

The situation becomes substantially more tricky
when primordial life forms do already exist
on the candidate planet, in a stage either
before or after the equivalent to the archean 
genetic expansion (which occurred on earth 
by 3.3-2.9\,Ga, see 
Sect.~\ref{sect:archean_genetic_expansion}).
The Genesis process could then lead to the 
destruction of a substantial fraction of 
indigenous lifeforms and therefore to a 
flagrant violation of the current consensus 
regarding planetary protection 
\citep{nicholson2009migrating}. In contrast one
may note that the microbes living on old
earth, being them bacteria or eukaryotes,
have never enjoyed human protection. Ethical or 
other type of arguments in favor of protecting
our terrestrial microbes are generically not 
voiced. Taking a deeper look one may argue that
planetary protection draws its justification 
from two sources \citep{lupisella2009search}:

\begin{itemize}
\item[--] The scientific benefit for humanity.
          Contaminating Mars or any other planet
          of the solar system with terrestrial
          microbes could ruin the possibility 
          to study non-terrestrial lifeforms.
          This argument does not apply to Genesis 
          candidate planets, which are selected 
          expressively for being out or range
          for in depth science missions. Planetary 
          protection will also break down, by the way, 
          once the doors of a manned spaceship opens
          on Mars \citep{campion2016moral}.
\item[--] Independently evolved life constitutes a value 
          per se \citep{randolph2014protecting}. This
          argument is actually closely related to core 
          motivation of the Genesis project, 
          namely that life as such is valuable.
\end{itemize}
It then boils down to the balancing of two options 
involving the prospect that the habitable lifespan
of the host planet may be too short for the indigenous 
life to evolve complex life by itself - the very reason 
the planet has been chosen - whereas the further evolved
precambrian terrestrial life may be prepared to do so.

\subsection{Seeding with unicellular organisms}

The simplest design for the seeding process would
be the from-orbit delivery via nano-sized reentry 
capsules, which could be in turn ejected backwards at
high velocities, viz decelerated with respect 
to the orbital motion of the main Genesis probe,
by a compact railgun \citep{poniaev2015small}. A 
minimal heat shield would then be enough to protect
the content of the delivery capsules during the 
subsequent drop to the surface. 

Key to the success prospects of the Genesis mission
is capability of the probe to use a databank of 
terrestrial genomes for the selection of the right
mix of microbes to be synthesized in situ by the
on-board gene laboratory. This brew of prokaryotes 
(bacteria) and unicellular eukaryotes will be
optimized with respect to the requirements 
resulting from the geophysical conditions of the 
host planet. It could be advantageous for the
Genesis probe to spread the seeding process 
over several centuries, albeit with slowly 
adapting mixtures of microbes.

The goal of the Genesis mission is to fast forward
the target planet to a precambrian state (see 
Fig.~\ref{fig:earthHistory}). Life would be given 
a head start consisting of a biosphere of unicellular 
organisms, from which on it could further flourish and 
develop. Direct seeding with multicellular organisms wouldn't
be impossible per se, but both substantially more 
complex and in part also questionable.
\begin{itemize}
\item[--] A planet with substantial levels of 
      $\mathrm{O}_2$ may be expected to develop
      complex life on its own. Planetary
      protection arguments in conjunction with
      possible biosphere incompatibilities would
      then dictate not to bring higher life forms
      to its surface (nor to seed it in first place).
\item[--] A planet without $\mathrm{O}_2$, the most
      probable situation, could also be seeded with
      multicellular life, as a matter of principle,
      but only with fully anaerobic animals of
      the type that are thought to dwell in earth's 
      oxygen-free deep sea environments (their cells 
      are devoid of mitochondria, possessing however 
      hydrogenosomes \citep{mentel2010anaerobic}).
      It is however questionable whether a passively 
      dropping reentry probe could successfully deliver
      these sub-millimeter animals to their proper habitats.
\end{itemize}
Even though most eukaryotes are aerobic and hence dependent
on free oxygen, a wide range of unicellular eukaryotes
have adapted to anoxic environments \citep{muller2012biochemistry}.
Seeding of a planet devoid of free oxygen with eukaryotes 
will hence not pose a problem \citep{yu2012evolutionary}.

\subsection{Post-seeding evolution and the oxidation of the atmosphere}

The mission of the Genesis project is to lay the foundations
for a self-evolving biosphere. It is however clear that fine 
tuning won't be possible and that the primary seeding process 
will result at best in a highly unbalanced ecosystems of microbes.
Global scale `ecological disasters' are hence expected to occur 
initially in the post-seeding phase, such as the uncontrolled
blossoming of unicellular algae. The ecosystem should however
self-stabilize relatively fast, say within a few thousand years.
The further evolution will then depend on a flurry of parameters,
like the initial concentration of atmospheric $\mathrm{CO}_2$,
the average temperature, the eventual presence of continents and
the overall level of hydrothermal activity.

It is presently not possible to estimate reliably
how long it will take afterwards for the photosynthetically
produced O$_2$ to accumulate in the atmosphere
of a Genesis plant.
\begin{itemize}
\item[--] Nearly all excess oxygen ever produced by 
          earth's biosphere has been used oxidizing
          the crust, compare Fig.~\ref{fig_organic_carbon_cycle},
          with less than 1/34 accumulating in the end in
          the atmosphere.
\item[--] Planets may dispose of quite different crustal compositions 
          in terms of the relative percentages of oxygen and 
          reducing elements (see Table~\ref{table_elements}).
          \begin{itemize}
          \item Both the bare metallicity (the abundance of
                heavy elements) and the relative abundances
                of the heavy elements will be distinct to 
                a planetary system \citep{longstaff2014astrobiology}.
          \item The initial core-crust segregation will depend
                likewise on the then present conditions, like
                the overall mass of the planet and the amount of
                radioactive heating \citep{longstaff2014astrobiology}.
          \item The post-segregation deposition of elements by 
                comets and asteroids, see Sect.~\ref{sec:LTB},
                may also influence the composition of the crust.
          \end{itemize}
\item[--] Non-biological processes like
$$
  2\mathrm{FeO} + \mathrm{H}_2\mathrm{O}\ \to\ 
   \mathrm{Fe}_2\mathrm{O}_3 + \mathrm{H}_2
$$
          contribute additionally to the net oxidation of the 
          crust whenever the resulting H$_2$ molecule manages 
          to escape to space \citep{wallace2006atmospheric}.
\end{itemize}
About 60\% of all atoms present in the crust of the earth 
are oxygen atoms and one may wonder whether this percentage 
is already close to saturation, at least with respect of 
what may be achievable by geo-planetary processes. It 
would be in any case optimal if the oxidation of the crust 
through antecedenting inorganic processes would be in an
advanced stage. This could be expected to be the case
for planets with delayed habitability and with elevated
stratospheric $\mathrm{H}_2\mathrm{O}$ concentrations 
\citep{catling2001biogenic}, allowing in turn for hydrogen
to escape into space (see the analogeous discussion in
Sect.\,\ref{sec:hadean_CO_2}).

A constant and hopefully high flux of minerals and 
$\mathrm{CO}_2$ is a general precondition for a
Genesis planet to develop a high bioproductivity
and hence a prerequisite also for a potentially 
rapid raise of atmospheric oxygen levels. We note 
here that about $11 \times 10^{18}$ out of the 
$8700 \times 10^{18}$ mol of organically produced 
C per Ma are buried in continental sediments 
\citep{field1998primary,sleep2001carbon}, 
nowadays on earth, from where it is recycled through 
carbonate weathering within about 350\,Ma (compare
also Fig.~\ref{fig_organic_carbon_cycle}). With
about $37 \times 10^{18}$ mol O$_2$ present in
the atmosphere \citep{jacob1999introduction}
that would imply that an atmosphere worth of O$_2$ 
is produced via organic CO$_2$ fixation every
$37/11=3.4$\,Ma. 

The balance of the biotic oxygen production through
the weathering of continental carbon deposits 
occurring on earth will however not start immediately
on planets not yet disposing of extended carbon sediments. 
The pace at which the atmospheric O$_2$ level rises
then depends on the overall bioproductivity and on
the amount of oxygen lost to the crust. An 
appreciable level could be achieved relatively fast, 
within 10-100\,Ma, if comparatively small amounts of
oxygen would be lost, viz whenever the crust of the
Genesis planet would already be in an advanced state 
of oxidation. The initial surge of oxygen levels
would be rebalanced in this scenario only once
the weathering rates have caught up respectively.

Most habitable planets will probably take of the order
of a few Ga to acquire an oxygen bearing atmosphere,
if at all. We are however confident that substantially 
shorter time scales may be achievable, as discussed
above, under optimal conditions and that we will hence 
be able to find Genesis candidate planets for which an 
initial seeding would initiate a geo-evolutionary process 
leading to the subsequent emergence of complex and multicellular 
life. Our best case estimates however still exceed human 
planning horizons by many orders of magnitude, implying 
that the Genesis process is intrinsically unsuited
for the preparation of a barren planet for an 
eventual human colonization. 



\section{Conclusions}

Today's scientific environment is made up by a diverse 
mix of emerging and mature fields, characterized
respectively by swift and lackluster rates of progress
\citep{gros2012pushing}. The sluggish progress of 
traditional space launching technologies \citep{ragab2015launch}
contrast here, e.g., with the rapid advances in synthetic 
biology \citep{stano2013semi,caspi2014divided}. Transformative
concepts are hence critical for reigniting innovation in science 
and technology time and again. It has been proposed 
in this context\citep{benford2013starship}, that robotic 
interstellar missions of low-weight crafts accelerated 
by beams of directed energy will become realizable, both 
on technical grounds and financially, in the near future 
\citep{projectDragonfly,breakthroughinitiatives}. At the 
same time we are discovering that planetary habitability 
isn't an all-or-nothing feature characterizing exoplanets
\citep{gudel2014astrophysical}. Our galaxy is expected
in particular to teem with planets which are in part
habitable, but for which the clement conditions do not 
last long enough for higher life forms to evolve on their 
own. 

Reversing the argument we have pointed out in this study 
that complex life may emerge also on transiently habitable
exoplanets whenever the extraordinary long time it took earth 
to develop eukaryotic cells could be leapfrogged. We have
argued furthermore that this endeavor could be achieved by 
a light-weight interstellar craft using a robotic gene 
laboratory for the seeding the target exoplanet with a 
brew of in situ synthesized microbes. By the end of the 
mission, which we call the Genesis project, a precambrian 
and hopefully thriving biosphere of unicellular organisms
would flourish on the candidate planet. Complex life
in the form of multicellular animals and plants will
evolve autonomously at a later state once the 
photosynthetically produced oxygen has had the time 
to accumulate in the atmosphere.

One of the key issues remaining to be settled at 
this stage regards the selection procedure for target
planets. Remote sensing of exo-planetary biosignatures 
from earth is possible \citep{des2002remote}, albeit 
only to a certain degree. An even more daring task would 
be to actually prove that a world is uninhabited 
\citep{persson2014does}. It is hence clear that the
final decision to go ahead must be taken autonomously 
by the on-board artificial intelligence. This may seem 
an imprudent strategy nowadays, but possibly not so 
in a few decades.

The Genesis mission is furthermore unique in the sense
that the actual cruising velocity is of minor importance.
It could be launched with the help of suitable beams of 
directed energy and decelerated at arrival by time consuming 
passive means like magnetic sails. We hence believe that 
the Genesis project opens a new venue for interstellar
missions and for the unfolding of life in our galactic 
surroundings.


\bibliographystyle{spr-mp-nameyear-cnd}
\bibliography{grosGenesis_astr}
\end{document}